\documentclass{article}

\usepackage{amssymb,dsfont,latexsym}

\newtheorem{thm}{Theorem}

\newtheorem{lem}[thm]{Lemma}
\newtheorem{cor}[thm]{Corollary}

\newcommand\bp{\noindent{\it Proof.}\ }

\newcommand\id{{\rm id}}

\newcommand\rank{{\rm rank}\,}

\newcommand\Tr{{\rm Tr}}

\begin{document}

\author{Erling St{\o}rmer}

\date{22-9-2005 }

\title{A reduction theorem for capacity of positive maps}

\maketitle
\begin{abstract}
We prove a reduction theorem for capacity of positive maps of finite dimensional $C^*-$algebras, 
thus reducing the computation of capacity to the case when the image of a nonscalar
 projection is never a projection.
\end{abstract}

\section*{Introduction}
In quantum information theory there has been a great deal of
interest in the concept of capacity of completely positive maps. A
drawback with capacity is that it is usually quite difficult to
compute, hence there is a need for developing computational
techniques. In the present paper we shall prove a reduction
theorem for capacity which reduces its computation to the ergodic
case. As a consequence we get a partial result towards the
additivity of capacity for tensor products.

If $P$ is a finite dimensional $C^*-$algebra we denote by $\Tr_P$
the trace on $P$ which takes the value 1 at each minimal
projection. Let $\eta$ denote the real function $\eta(t)=-t\log t
$ for $t>0$, and  $ \eta(0)=0.$ Then the entropy $S(a)$ of a
positive operator $a$ in $P$ is defined by $S(a)=\Tr_P(\eta(a)).$
If $M$ is another finite dimensional $C^*-$algebra let $\Phi\colon
M\to P$ be a positive unital linear trace preserving map, i.e.
$\Tr_P(\Phi(x))=\Tr_M(x)$ for all $x\in M.$ Note that we only
assume $\Phi$ is positive and not completely positive, since the
latter stronger assumption is in most cases unnecessary. Let $C$
denote the positive operators in $M$ with trace 1.  If $a\in C$
let
$$
C(\Phi,a)= \sup S(\Phi(a)) - \sum_i \lambda_i S(\Phi(a_i)),
$$
where the sup is over all convex combinations of operators $a_i
\in C$ with $\sum_i \lambda_i a_i =a.$ The \textit{capacity}
$C(\Phi)$ of $\Phi$ is defined by
$$
C(\Phi)=\sup_{a\in C}C(\Phi,a).
$$
For a discussion of capacity see e.g.  \cite{H}.

\section{The reduction theorem}
If $P$ is a finite dimensional $C^*-$algebra and $\omega$ is a
state on $P$ let $Q_\omega$ denote its density operator in $P.$
Then the entropy of $\omega$ (with respect to $P$) is
$S(\omega)=S(Q_\omega).$ We shall need three properties of
entropy, namely: it is subadditive, i.e. $S(\omega_1 +
\omega_2)\leq S(\omega_1)+S(\omega_2);$ it is concave, i.e.
$S(\lambda \omega_1 + (1-\lambda)\omega_2)\geq \lambda
S(\omega_1)+ (1-\lambda)S(\omega_2),$ and if $N\subseteq
M\subseteq P$ are $C^*-$subalgebras then $S(\omega\mid N)\geq
S(\omega\mid M).$ Our first result is taken from the book  \cite{NS} and is 
an inequality in the
opposite direction.
\begin{lem}\label{lem}
Let $M\subseteq P$ be finite dimensional $C^*-$algebras, and let
$e_1,\dots,e_n$ be projections in $M$ with sum 1.  Let
$N=\bigoplus_{i=1}^n N_i,$ where $N_i=e_i Me_i.$ Let $\omega$ be a
state on $P.$ Then
$$
\sum_i
\omega(e_i)S(\frac{\omega|N_i}{\omega(e_i)})=S(\omega|N)-\sum_i
\eta(\omega(e_i))\leq S(\omega).
$$
\end{lem}
\bp Let $s_i=\omega(e_i).$ Then
\begin{eqnarray*}
S(\omega|N)&=&\sum_i S(\omega(e_i . e_i))\\
 &=&\sum_i S(\frac{\omega(e_i . e_i)}{s_i}s_i)\\
 &=&\sum_i s_i S(\frac{\omega(e_i . e_i)}{s_i})+\eta(s_i)
\end{eqnarray*}
 which proves the equality in the lemma.

 In order to prove the inequality let $f_k$ be minimal
 projections in $P$ and $\alpha_k >0$ such that the density
 operator $Q_\omega$ for $\omega$ is of the form $Q_\omega =\sum_k \alpha_k
 f_k,$ so in particular $\sum_k \alpha_k=1.$ Thus $S(\omega)=S(Q_\omega)=\sum_k
 \eta(\alpha_k).$ By the first part of the proof we have
\begin{eqnarray*}
S(\omega|N)
&=&\sum_i S(\omega(e_i . e_i))\\
&=& \sum_i S(\sum_k \alpha_k e_i f_k e_i)\\
&\leq&\sum_{i,k}S( \alpha_k e_i f_k e_i)\\
&=&\sum_{i,k} \alpha_k S(e_i f_k e_i)+\eta(\alpha_k)\Tr_P (e_i f_k e_i)\\
&=&\sum_{i,k} \alpha_k \eta(\Tr_P(e_i f_k
e_i))+\eta(\alpha_k)\Tr_P (e_i f_k
e_i)\\
&\leq& \sum_i \eta(\sum_k \alpha_k\Tr_P(e_i f_k e_i))+\sum_k \eta(\alpha_k)\\
&=&\sum_i \eta(\Tr_P(e_i Q_\omega e_i)) + S(\omega)\\
&=&\sum_i \eta(\omega(e_i)) +S(\omega),
\end{eqnarray*}
where the first inequality follows from subadditivity of $S$ and
second from concavity. We also used that $e_i f_k e_i=Tr_P(e_i f_k
e_i)p$, where $p$ is a minimal projection. The proof is complete.

From the definition of capacity it is clear that if $\Phi\colon
M\to P$ is as before, and $N\subseteq M,$ then $C(\Phi|N)\leq
C(\Phi).$ Our next result describes a situation when we have
equality. We shall use a result of Broise, see \cite{S} , that if $a$ is
a self-adjoint operator in $M$ such that $\Phi(a^2)=\Phi(a)^2$
then $\Phi(aba)=\Phi(a)\Phi(b)\Phi(a)$ for all $b\in M.$ In
particular, if $e$ is a projection in $M$ such that $\Phi(e)$ is a
projection, then the above identity holds for $a$ replaced by $e.$
The ergodic case alluded to in the introduction is the case when
the only operators $a$ which satisfy $\Phi(a^2)=\Phi(a)^2$ are the
scalar operators.

\begin{thm}\label{thm1}
Let $M, P$ be finite dimensional $C^*-$algebras. Let $\Phi\colon
M\to P$ be a positive unital trace preserving map. Suppose
$e_1,\dots,e_n$ are projections in $M$ with sum 1 such that
$\Phi(e_i)$ is a projection for all $i$. Let $N=\bigoplus e_i
Me_i.$ Then $C(\Phi)=C(\Phi|N).$
\end{thm}
\bp Clearly $C(\Phi)\geq C(\Phi|N).$ For the opposite inequality
let $a, a_m \in C$ such that $a=\sum_m \lambda_m a_m.$ Let
$Q=\bigoplus \Phi(e_i)P\Phi(e_i).$ Since $\Phi(e_i
xe_i)=\Phi(e_i)\Phi(x)\Phi(e_i)$ for all $x\in
M,\Phi(E_N(x))=E_Q(\Phi(x))$, where $E_N$ and $E_Q$ denote the
conditional expectations on $N$ and $Q$ respectively. Thus
$$
S(\Phi(a))\leq S(E_Q(\Phi(a)))=S(\Phi(E_N (a))).
$$
Therefore by  Lemma 1 applied to the states $\omega_m$ defined by
$Q_{\omega_m}=\Phi(a_m)$ and $e_1,\dots,e_n$ yields the following
inequality.
\begin{eqnarray*}
& &S(\Phi(a))- \sum_m \lambda_m S(\Phi(a_m))\\ &\leq&
 S(\Phi(E_N(a)))-\sum_m \lambda_m\sum_i
\Tr_P(\Phi(e_i)\Phi(a_m)\Phi(e_i))S(\frac{\Phi(e_i)\Phi( a_m
)\Phi(e_i)}{\Tr_P(\Phi(e_i)\Phi(a_m)\Phi(e_i))})\\
&=& S(\Phi(E_N(a)))-\sum_m \lambda_m\sum_i\Tr_P(\Phi(e_i a_m
e_i))S(\frac{\Phi(e_i a_m e_i)}{\Tr_P(\Phi(e_i a_m e_i)})\\
&=&S(\Phi(E_N(a))) - \sum_{m,i} \lambda_m\Tr_M(e_i a_m
e_i)S(\frac{\Phi(e_i a_m e_i)}{\Tr_M(e_i a_m e_i)})\\
&=&S(\Phi(E_N(a))) -\sum_{m,i}\mu_{m,i}S(\frac{\Phi(e_i a_m
e_i)}{\Tr_M(e_i a_m e_i)}),
\end{eqnarray*}
where $\sum_{m,i}\mu_{m,i}=1,$ and $\frac{e_i a_m e_i}{\Tr_M(e_i
a_m e_i)}=E_N(\frac{e_i a_m e_i}{\Tr_M(e_ia_m e_i)})\in N$ with
trace 1. Since the above inequality holds for all families $(a_m)$
as above
$$
C(\Phi,a)\leq C(\Phi|N,E_N(a)).
$$
Since this holds for all $a\in M$
$$
C(\Phi)=\sup_a C(\Phi,a)\leq \sup_a C(\Phi|N,E_N(a))=C(\Phi|N),
$$
proving the theorem.

 We can now state our main reduction theorem.
Note that if the projections $e_i$ are minimal with the property
that $\Phi(e_i)$ is a projection, then $\Phi|e_i Me_i$ is ergodic
in the sense defined above, so the theorem is a reduction to the
ergodic case.

\begin{thm}\label{thm2}
Let $M, P$ be finite dimensional $C^*-$algebras and $\Phi\colon
M\to P$ a positive unital trace preserving map. Let
$e_1,\dots,e_n$ be projections in $M$ with sum 1 such that $\Phi(e_i)$ is a
projection for each i. Let $M_i=e_i Me_i$ and $\Phi_i =\Phi|M_{i}
\colon M_{i} \to \Phi(e_i)P\Phi(e_i)$ be the restriction map to
$M_i.$ Then
$$
C(\Phi)=\log \sum_{i=1}^n e^{C(\Phi_i)}.
$$
\end{thm}
\bp By Theorem 2 it suffices to consider $a=\sum_i a_i \in M,
a_i=ae_i \in M_i,$ where $a_i=\sum_j \lambda_{ji}a_{ji}$ with
$\Tr_M(a_{ji})=1$, $a_{ji}\in M_{i}^+,$ $\sum_{ji}\lambda_{ji}=1.$
Let $s_i=\Tr_M(e_i a)=\Tr_M(a_i)=\Tr_P(\Phi(e_i)\Phi(a)).$ Then we
have
\begin{eqnarray*}
S(\Phi(a))&-&\sum_{ji}\lambda_{ji}S(\Phi(a_{ji}))\\ &=&
\sum_i[S(\Phi(e_i)\Phi(a))-\sum_j\lambda_{ji}S(\Phi(a_{ji}))]\\
&=&\sum_i[S(s_i(\frac{1}{s_i}\Phi(e_i)\Phi(a)))-s_i\sum_j \frac{\lambda_{ji}}{s_i}S(\Phi(a_{ji}))]\\
&=&-\sum_i s_i\log s_i +\sum_i
s_i[S(\frac{1}{s_i}\Phi(e_i)\Phi(a))-\sum_j\frac{\lambda_{ji}}{s_i}S(\Phi(a_{ji}))]
\end{eqnarray*}
We have
$$
S(\frac{1}{s_i}\Phi(e_i)\Phi(a))-\sum_j\frac{\lambda_{ji}}{s_i}S(\Phi(a_{ji}))\leq
C(\Phi|M_i).
$$
Therefore
\begin{eqnarray*}
S(\Phi(a))&-&\sum_{ji}\lambda_{ji}S(\Phi(a_{ji}))\\
&\leq& -\sum_i s_i(\log s_i - C( \Phi|M_i))\\
&=&-\sum_i s_i(\log s_i - \log\frac{C(\Phi|M_i)}{\sum_k
e^{C(\Phi|M_k)} })+\log \sum_i e^{C(\Phi|M_i)}
\end{eqnarray*}
Since the sum $\sum_i s_i(\log s_i -
\log\frac{e^{C(\Phi|M_i)}}{\sum_k e^{C(\Phi|M_k)} }) $ is a
relative entropy, it is nonnegative, see Lemma 4.5 in ~\cite{Sm}.
Hence we have
$$
S(\Phi(a))-\sum_{ji}\lambda_{ji}S(\Phi(a_{ji}))\leq \log \sum_i
e^{C(\Phi|M_i)},
$$
Since this holds for all $a$ we conclude that $C(\Phi)\leq\log
\sum_i e^{C(\Phi|M_i)}.$

For the converse inequality let $\varepsilon >0,$ and choose $b_i
\in M_i^+$ with $\Tr_M(b_i)=1,$ $\mu_{ji}\geq 0$ with $\sum_{j} \mu_{ji}=1$ and
$a_{ji}\in M_i^+$ with trace 1 such that $\sum_{j} \mu_{ji}
a_{ji}=b_i,$ and
$$
S(\Phi(b_i))-\sum_j \mu_{ji}S(\Phi(a_{ji}))\geq
C(\Phi|M_i)-\varepsilon.
$$
Let now $s_i\geq 0$ have sum 1, and let $a_i= s_i b_i,$
$\lambda_{ji}=s_i \mu_{ji}.$ Put $a=\sum_i a_i
=\sum_{ji}\lambda_{ji}a_{ji}.$ Then by the above inequality we
have
$$
S(\frac{1}{s_i}\Phi(e_i)\Phi(a_{i}))-\sum_j\frac{\lambda_{ji}}{s_i}S(\Phi(a_{ji}))\geq
C(\Phi|M_i)-\varepsilon.
$$
Thus by the computations in the beginning of the proof we have
$$
S(\Phi(a))-\sum_{ji}\lambda_{ji}S(\Phi(a_{ji}))\geq -\sum_i
s_i(\log s_i - C(\Phi|M_i)) -\varepsilon.
$$
Hence by the same computation we did above we obtain
\begin {eqnarray*}
S(\Phi(a))&-&\sum_{ji}\lambda_{ji}S(\Phi(a_{ji}))\\
&\geq& -\sum_i
s_i(\log s_i - \log{\frac{C(\Phi|M_i)}{\sum_k e^{C(\Phi|M_k)}}
})+\log \sum_k e^{C(\Phi|M_k)}-\varepsilon.
\end{eqnarray*}
For the value $s_i=\frac{C(\Phi|M_i)}{\sum_k C(\Phi|M_k)}$ the
value of the relative entropy is 0, hence
$$
C(\Phi)\geq
S(\Phi(a))-\sum_{ji}\lambda_{ji}S(\Phi(a_{ji}))\geq\log \sum_k
e^{C(\Phi|M_k)}-\varepsilon.
$$
Since $\varepsilon$ is arbitrary the proof is complete.

A good illustration of an application of the theorem is the case
when $\Phi$ is a trace preserving projection map of $M$ into
itself, i.e.$ \Phi(x)=\Phi(\Phi(x))$ for all $x\in M.$ Then the
image $N=\Phi(M)$ is a Jordan subalgebra of $M,$ and if $\Phi$ is
completely positive then $\Phi$ is a conditional expectation, and
$\Phi(M)$ is a $C^*-$algebra, see ~\cite{ES}. The rank of $N$
-$rank N$- is the maximal number of minimal projections in $N$
with sum 1.

\begin{cor}\label{cor1}
Let $\Phi\colon M\to M$ be a trace preserving projection map. Then
$$
C(\Phi)=\log \rank \Phi(M).
$$
\end{cor}
\bp Let $n=\rank N$ and $e_1,\dots,e_n$ be minimal projections in
$\Phi(M)$ with sum 1. Then $e_k Me_k =\0C e_k$ for all $k,$ hence
$C(\Phi|e_k Me_k)=0,$ so by the theorem
$$
C(\Phi)=\log\sum_i ^n e^0 = \log n.
$$
The proof is complete.

The main problem concerning capacity is whether it is additive
under tensor products, i.e. whether
$C(\Phi\otimes\Psi)=C(\Phi)+C(\Psi)$ when $\Phi\otimes\Psi$ is
positive, in particular when they are both completely positive.
Our next result reduces the problem to the case when both maps are
ergodic.

\begin{cor}\label{cor2}
Let $M,N,P,Q$ be finite dimensional $C^*-$algebras and $\Phi\colon
M\to P$ and $\Psi\colon N\to Q$ be positive unital trace
preserving maps such that $\Phi\otimes\Psi\colon M\otimes N \to
P\otimes Q$ is positive. Let $e_i\in M$ and $f_j\in N$ be
projections with sum 1 such that $\Phi(e_i)$ and $\Psi(f_j)$ are
projections. Let
$$
\Phi_i =\Phi|e_i Me_i \colon e_i Me_i \to \Phi(e_i)P\Phi(e_i),
$$
$$
\Psi_j =\Psi|f_j Nf_j \colon f_j Nf_j \to \Psi(f_j)Q\Psi(f_j).
$$
Suppose $C(\Phi_i \otimes \Psi_j)=C(\Phi_i)+C(\Psi_j)$ for all
$i,j.$ Then
$$
C(\Phi\otimes\Psi)=C(\Phi)+C(\Psi).
$$
\end{cor}
\bp We apply Theorem 3 to the projections $e_i\otimes f_j$ and the
corresponding maps $\Phi_i \otimes \Psi_j$. Thus we have
\begin{eqnarray*}
C(\Phi\otimes\Psi) &=& = \log \sum_{ij}e^{C(\Phi_i \otimes \Psi_j})=\log\sum_{ij} e^{C(\Phi_i)+C(\Psi_j)}\\
&=&\log\sum_{ij} e^{C(\Phi_i)}  e^{C(\Psi_j)}=
\log\sum_{i}e^{C(\Phi_i)}\sum_{j}e^{C(\Psi_j)}\\
&=&C(\Phi)+C(\Psi). \end{eqnarray*} The proof is complete.

If $\Phi$ is completely positive and $id$ is the identity map of
$N$ let $f_j$ be a minimal projection for each $j.$ Then the
assumptions of the above corollary hold for the projections
$1\otimes f_j.$ Hence we have

\begin{cor}\label{cor3}
Let $M$ and $N$ be finite dimensional $C^*-$algebras as before
with $\Phi$ completely positive. Then $C(\Phi\otimes\id)=
C(\Phi)+\log rank N.$
 \end{cor}

Department of Mathematics, University of Oslo, 0316 Oslo, Norway.

e-mail erlings@math.uio.no

\end{document}